\documentstyle[prl,aps,epsf]{revtex}
\begin{document}
\title{The monoclinic phase of PZT ceramics: Raman and phenomenological theory
studies}
\author{A. G. Souza Filho\thanks{
Corresponding Author: A. G. Souza Filho, FAX 55 (85) 2874138,
E-mail: agsf@fisica.ufc.br}, K. C. V. Lima, A. P. Ayala, I.
Guedes, P. T. C. Freire and J. Mendes Filho}
\address{Departamento de F\'{\i}sica, Universidade Federal do
Cear\'{a}, Caixa Postal 6030,\\ Campus do Pici, 60455-760
Fortaleza, Cear\'{a}, Brazil}
\author{E. B. Ara\'ujo and J. A. Eiras}
\address{Departamento de F\'{\i}sica, Universidade Federal de S\~ao Carlos,
Caixa Postal 676,\\ 13565-670 S\~ao Carlos SP - Brazil}
\date{\today}
\maketitle

\begin{abstract}
This work reports on the first Raman detection of the tetragonal
$\rightarrow $ monoclinic phase transition in PZT ceramics near
morphotropic phase boundary at low temperatures. The transition is
characterized by changes in the frequency of lattice modes with
the temperature. The results presented here confirm the previous
one recently reported by Noheda \textit{et al.}\cite{bn,jag} using
high-resolution synchrotron X-ray powder diffraction technique and
dielectric measurements. The stability of the new phase is
discussed within the framework of phenomenological
Landau-Devonshire Theory.
\end{abstract}


Due to their remarkable technological importance,
PbZr$_{1-x}$Ti$_{x}$O$_3$ ceramics have been for long time the
subject of extensive experimental and theoretical studies. Its
excellent piezoelectric, dielectric and ferroelectric properties
lead this material to be one of the most important material used
in electronic devices. Its properties depend strongly on its
structural phase. The composition($x$)-temperature(T) phase
diagram of the PZT system was determined by Jaffe \textit{et
al.}\cite{bj} using X-ray diffraction measurements. At room
temperature PZT presents several structures  depending on the
value of $x$, namely: ferroelectric tetragonal (C$_{4v}^1$), two
rhombohedral phases (C$_{3v}^6$ and C$_{3v}^1$), and
antiferroelectric orthorhombic (C$_{2v}^8$) phases. Also, at high
temperatures, PZT presents a paraelectric cubic phase (O$_h^1$).
The phenomenological Landau-Devonshire theory developed by Haun
\textit{et al.} \cite{hau89a} explains the stability of all phases
contained in the concentration($x$)-temperature(T) phase diagram
determined by Jaffe \textit{et al.}\cite{bj}.

The separation between rhombohedral and tetragonal phase is called
morphotropic phase boundary (MPB) and occurs around x = 0.50 and
0.55, where the exact compositions which the transitions occur are
strongly dependent on the sample fabrication method \cite{kk} and
the grain sizes \cite{rsk}. Many efforts have been concentrated in
obtaining samples belonging to this region of phase diagram,
because of the enhancement of their electromechanical response.
So, recently, new features on the MPB were reported by Noheda et.
al \cite {bn,jag}. Their works report on the high-resolution
synchrotron X-ray powder diffraction techniques and dielectric
measurements in PZT ceramics with compositions slightly different
from that of MPB region. The authors discovered a new
ferroelectric monoclinic phase at low temperatures and proposed a
preliminary modification in the well-known phase diagram reported
by Jaffe et. al. \cite{bj}. The most probable space group for this
new ferroelectric monoclinic phase is C$_s^3$ which is a subgroup
of C$_{4v}^1 $ and C$_{3v}^6$ groups. This monoclinic phase
exhibited by PZT at low temperatures would be the first example of
a ferroelectric material with P$_x^2$=P$_y^2$$\neq $P$_z^2$, where
P$_x^2$, P$_y^2$, P$_z^2$ $\neq$ 0. \cite{bn}

The transition observed by Noheda \textit{et al.}\cite{bn} was due
to the use of synchrotron radiation which provides a
higher-resolution diffraction pattern than that obtained by
conventional X-ray radiation. Micro-Raman spectroscopy, which is
quite useful to investigate a localized area in the probed sample
with a spatial resolution of the order of $\mu$m, is very
effective tool to study phase transitions in polycrystalline
materials. Despite the works reported in Refs. [5] and [6] no
study on the modifications of optical modes when PZT undergoes a
phase transition from tetragonal to monoclinic were already
presented. So, the aim of this work is to present a systematic
micro-Raman study of PZT ceramics close to MPB in the temperature
range from 25 to -240 $^oC$, in order to reveal the behavior of
optical phonons when PZT ceramics undergoes a phase transition
from tetragonal to monoclinic structure. A preliminar analysis of
the phase sequence, based on phenomenological Landau-Denvonshire
theory, is also presented.

The sample PbZr$_{0.52}$Ti$_{0.48}$O$_3$ was obtained through the
solid-state reaction from 99.9 $\%$ pure reagent grade PbO, ZrO$_2$ and TiO$%
_2$ oxides. The starting powders and distilled water were mixed
and milled during 3.5 hours for powder homogenization. The mixture was calcined at 850 $%
^oC$ for 2.5 h. It was pressed at 400\thinspace MPa giving rise to
PZT ceramics disks with 10\thinspace mm of diameter and
5\thinspace mm of thickness. Finally, the disks were sintered at
temperature of 1250 $^oC$ during 4h. The sintering atmosphere,
which consists of a covered alumina crucible, was enriched in PbO
vapor using PbZrO$_3$ powder around the disks in order to avoid
significant volatilization of PbO.

MicroRaman measurements were performed using a T64000 Jobin Yvon
Spectrometer equipped with an Olympus microsocope and a N$_2$-
cooled CCD to
detect scattered light. The spectra were excited with an Argon ion Laser ($%
\lambda$ = 514.5 nm), and the power laser impinging on the samples
surface is estimated to be of order of 10 mW. The spectrometer
slits were set to give a spectral resolution of better than 2
cm$^{-1}$. A Nikon 20x objective with focal distance 20 mm and
numeric apperture N.A. = 0.35 was employed to focus the laser beam
on the polished sample surface. Low temperature measurements were
performed using an Air Products closed-cycle refrigeration which
provides temperatures ranging from 20 K to 300 K. An Eurotherm
controller was used to control the temperature with precision of order of $%
\pm$0.1 K.

Raman spectra in the low frequency mode region of PbZr$_{0.52}$Ti$_{0.48}$O$%
_3$ ceramic recorded at several temperatures are presented in Fig.
1. First, let us discuss on the assignment of the Raman modes
observed in the spectrum at room temperature. The spectral region
below 150 cm$^{-1}$ contains the external modes of the material
and for PZT systems is related to Pb- lattice modes\cite{car92},
while the region above 150 cm$^{-1}$ contains the modes related to
some internal modes (bending, torsion, and stretching) of certain
groups of the material, e.g., (Zr/Ti)O$_6$ units. The Raman modes
located at 25, 48, 94, and 136 cm$^{-1}$ are labeled as A, B, C,
and D, respectively. The mode at 50 cm$^{-1}$ , which is absent in
PbTiO$_3$, appears in bulk PZT ceramics for Zr content lower than
0.1, being considered an additional mode \cite{rsk,rm}. In our
samples, this mode were found to be located at 48 cm$^{-1}$.

\begin{figure}[htbp]
\protect \centerline {\epsfxsize=4.1in \epsfbox{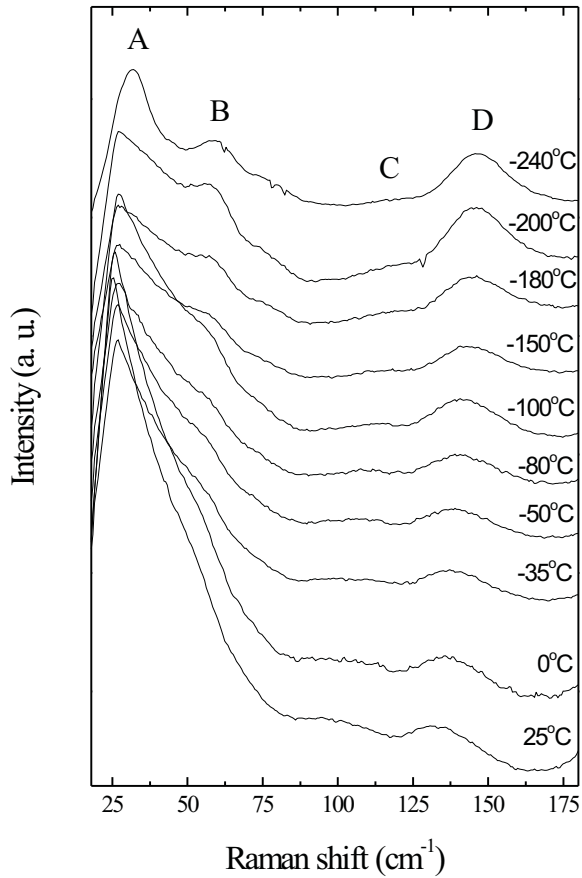}}
\caption{Low frequency Raman spectra of
PbZr$_{0.52}$Ti$_{0.48}$O$_3$ ceramics recorded in the temperature
range from 25 to -240 $^oC$.}\label{fig1}
\end{figure}

Next, let us discuss on the temperature dependence of the
frequency modes shown in Fig. 1. Particular attention was devoted
to the modes A, B, and C. The mode B, which at room temperature
appears to be coalesced with mode A, is clearly solved at low
temperatures. In the temperature interval from -10 to -40 $^oC$,
it was observed changes in the frequency of mode C, which are
depicted in Fig. 2. Such kind of behavior clearly indicates the
existence of a phase transition, which following the works
reported by Noheda \textit{et al.}\cite{bn,jag}, changes the
structure of PZT from tetragonal (C$_{4v}^1$) to monoclinic
(C$_s^3$).

\begin{figure}[htbp]
\protect \centerline {\epsfxsize=4.0in \epsfbox{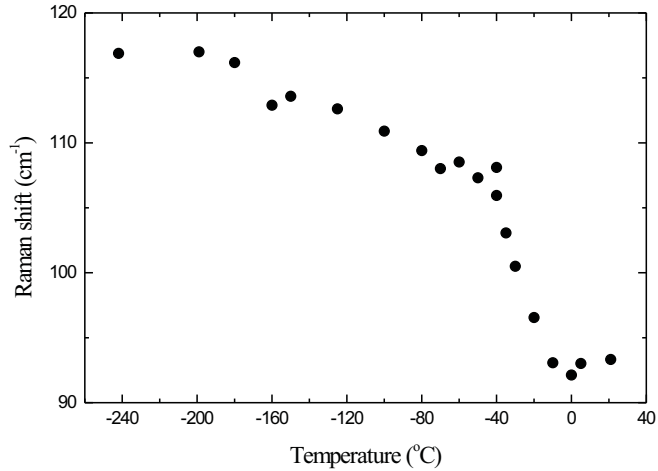}}
\caption{Variation of the frequency of the Raman mode C as a
function of temperature in the range from 25 to -240 $^oC$. The
anomaly observed clearly indicates the occurrence of a phase
transition from tetragonal to monoclinic structure.}\label{fig2}
\end{figure}

In order to discuss the stability of this new phase we use the Landau$-$%
Devonshire phenomenological theory. It should be remarked that,
Haun and coworkers \cite {hau89a,hau89b,hau89c,hau89d,hau89e} have
evaluated various coefficients related to the free energy
function, developing a complete theory that describes the
well-known phase transitions in the PZT system. In this work, we
consider a simplified version of the free energy used in Ref.[2],
since only polarization changes are involved in the phases
separated by the MPB. In this case, the free energy, in absence of
stress, is given by :

\begin{eqnarray}
\nonumber \Delta G = \alpha_1 ( P_1^2 + P_2^2 + P_3^2 ) +
\alpha_{11} ( P_1^4 + P_2^4 + P_3^4 )+ \alpha _{12} (P_1^2\,P_2^2
+ P_1^2\,P_3^2 + P_2^2\,P_3^2 )+\\ + \alpha_{111} ( P_1^6 + P_2^6
+ P_3^6 ) + \alpha _{112} (P_3^4 ( P_1^2 + P_2^2 ) + P_2^4\, (
P_1^2 + P_3^2 ) +\\ \nonumber+ P_1^4\, ( P_2^2 + P_3^2 ) )+ \alpha
_{123} P_1^2\,P_2^2\,P_3^2
\end{eqnarray}

\noindent where $P_i$ (i=1,2,3) is the polarization along the
cubic axes, $ \alpha_1, \alpha_{i,j}$ (i, j = 1,2), and
$\alpha_{i,j,k}$ (i,j,k=1,2,3) are the dielectric stiffness and
high-order dielectric stiffness at constant stress, respectively.
Here $\alpha_1$ is the only temperature dependent coefficient and
it is given by the Curie-Weiss law ( $\alpha_1 = (T-T_0)/C
\epsilon_0$). As already mentioned, the values of all coefficients
and the solutions for the tetragonal and rhombohedric phases can
be found in Refs. [2] and [9-12]. Here, we present only the
solutions for the monoclinic phase as proposed by Noheda
\textit{et al.}\cite{bn}. These authors stated that the cations
displacements lie close to the monoclinic [$\bar 2$01] direction,
which corresponds to the rhombohedral [111] axis. As a consequence
of this structure, the electric polarization
have the following conditions: P$_1^2$=P$_2^2$$\neq $P$_3^2$ and P$_1^2$, P$%
_3^2$ $\neq$ 0. In this case, the free energy under zero stress
reads

\begin{eqnarray}
\nonumber \Delta G_{FM} = \alpha_1(2 P_1^2 + P_3^2) + (2
\alpha_{11}+ \alpha_{12}) P_1^4 + \alpha_{11} P_3^4 + 2
\alpha_{12} P_1^2 P_3^2  + 2 (\alpha_{111}+\alpha_{112}) P_1^6
\nonumber +\\+ \alpha_{111} P_3^6 + 2 \alpha_{112} P_1^2 P_3^4 +
(2 \alpha_{112}+\alpha_{123}) P_1^4 P_3^2
\end{eqnarray}

As the free energy of the monoclinic phase has two order
parameters, the spontaneous polarization can be found from the
first partial derivative stability conditions:
$\partial{\Delta G_{FM}}/{\partial P_1}=0$ and $\partial {\Delta G_{FM}}/ {%
\partial P_3}=0$. Unfortunately, this is a system of nonlinear equations
which does not have an analytical solution. Hence a numerical
solution method was used to obtain the polarization and the free
energy of the monoclinic phase. In Fig. 3 the free energy of the
tetragonal, rhombohedric and monoclinic phases is plotted as a
function of the Ti concentration at -80 $^oC$. While Eq. [2] is
able to predict the rhombohedric $\rightarrow$ tetragonal phase
transition at approximately $x$=0.5, the calculations clearly
shows that the monoclinc phase is not stable for any value of Ti
concentration. This result indicates that Eq. (2) is not suitable
to describe the existence of monoclinic phase. To describe the
monoclinic phase we take into account a third order parameter,
which in accordance with Ref. 6, is the monoclinic angle $\beta$.
In fact, the monoclinic angle $\beta$ is not the true order
parameter since it does not go to zero at T=T$_M$, where T$_M$ is
the temperature of tetragonal $\rightarrow$ monoclinic phase
transition. The true order parameter should be $90^o-\beta$.
However, we will maintain the notation $\beta$ for the third order
parameter. With this, the PZT system in the monoclinic phase has
the characteristic of an improper ferroelectric \cite{str97} and
the free energy is now given by:

\begin{figure}[htbp]
\protect \centerline {\epsfxsize=4.2in \epsfbox{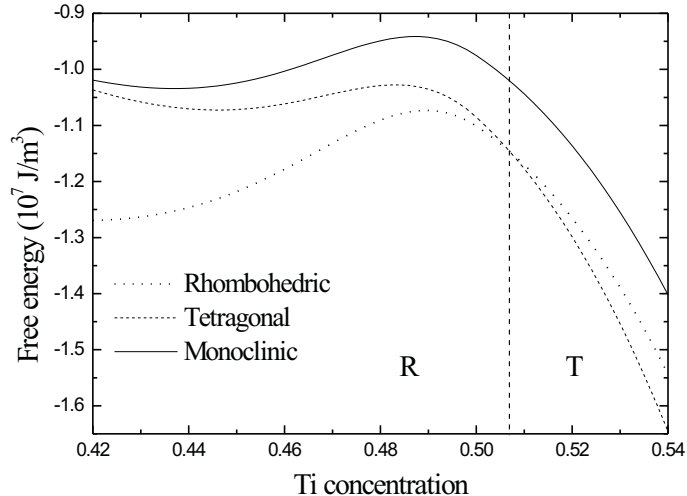}}
\caption{Plot of free energy as a function of Ti concentration for
the rhombohedric (R), tetragonal (T) and monoclinic (M) solutions
at T=-80$^oC$ . The capital letters indicated the low energy phase
in each region.}\label{fig3}
\end{figure}

\begin{equation}
\nonumber \Delta G_{IM} = \Delta G_{FM} + \eta_1 (T-T_M) \beta^2 +
\eta_{11} \beta^4 + \phi \beta^2 (2 P_1^2 + P_3^2)
\end{equation}

Unfortunately, there is no much information about the new
monoclinic phase. So we do not have the experimental data
necessary to calculate the values of the free energy coefficients.
However, to verify that Eq. (3) is able to predict the existence
of the monoclinic phase, we will make some rough approximations to
estimate these coefficients. First, we will suppose that the
coupling coefficient between $\beta$ and the polarization ($\phi$)
is negligible compared with the dielectric stiffness coefficient
($\alpha_1$). With this supposition, the first partial derivative
stability condition read

\begin{equation}
\partial{\Delta G_{IM}}/{\partial P_1} \simeq \partial{\Delta
G_{FM}}/{\partial P_1}=0
\end{equation}

\begin{equation}
\partial{\Delta G_{IM}}/ {\partial P_3} \simeq \partial {\Delta
G_{FM}} / {\partial P_3}=0
\end{equation}

\begin{equation}
\nonumber \partial{\Delta G_{IM}}/{\partial \beta} = 2\beta
(\eta_1(T-T_M) + \phi (2 P_1^2+P_3^2)+2 \eta_{11} \beta^2 )=0
\end{equation}

Thus, the polarization values obtained by minimizing the Eq. 2 can
be used to calculate the third order parameter, that is

\begin{equation}
\beta^2 = - \frac {\eta_1} {2 \eta_{11}}(T-T_M+ \frac{\phi}
{\eta_1}(2 P_1^2+P_3^2))
\end{equation}

Replacing Eq. (7) into Eq. (3), we find the modified free energy
in the monoclinic phase, namely

\begin{equation}\Delta G_{IM} = \Delta G_{FM} - \frac{\eta_1^2} {4
\eta_{11}} (T-T_M+ \frac{\phi} {\eta_1}(2 P_1^2+P_3^2))^2
\end{equation}

Next, we can use the few available data to estimate the unknown
coefficients. By using the temperature dependence of $\beta$
obtained by Noheda \textit{et al.} \cite{jag} and Eq. (7), we
calculated that the $\phi / \eta_1$ ratio is $\approx$ 10$^{3}$
m$^2$$^o$C/C. Furthermore, we assumed a linear dependence of the
transition temperature between the values obtained for Ti
concentrations between 0.48 and 0.50. Finally, the ${\eta_1^2}/ {4
\eta_{11}}$ ratio necessary for the Eq. 8 could be approximately
found by considering that the tetragonal and monoclinic free
energies should be equal on the phase transition ($x$ $\simeq$
0.48 and T$_M$ $\simeq$ -40 $^oC$; $x$ $\simeq$ 0.50 and T$_M$
$\simeq$ -80 $^oC$). This calculation yielded an approximate value
of ${\eta_1^2}/ {4 \eta_{11}}$ $\simeq$ 5 J/$^o$C$^2$m$^3$. By
using this estimated parameter, we were able to calculate the
dependence concentration of the modified free energy. In Fig. 4 we
compare the modified free energy with those of other phases of
PZT. It should be observed that the value of the free energy of
the monoclinic phase is lower than that of free energy of the
rhombohedric and tetragonal phases in a small region close to MPB.
This result shows that the modified free energy correctly predicts
the existence of a monoclinic phase in the MPB region, in
accordance with the experimental observations.

\begin{figure}[htbp]
\protect \centerline {\epsfxsize=4.2in \epsfbox{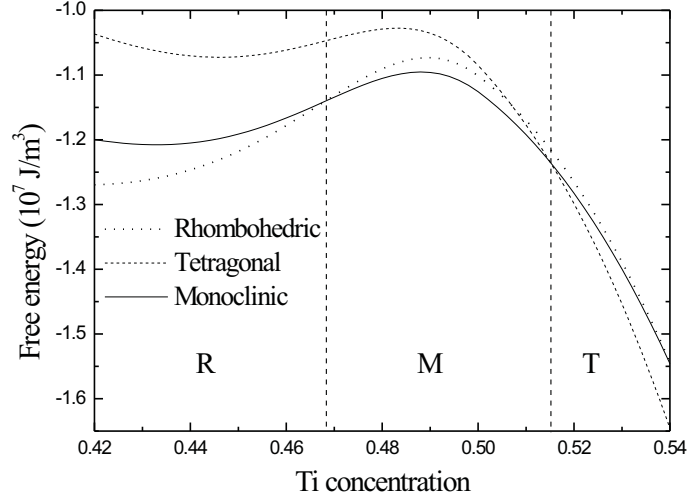}}
\caption{Plot of free energy as a function of Ti concentration for
the rhombohedric (R), tetragonal (T) and monoclinic (M) solutions
at T=-80$^oC$ . The capital letters indicated the low energy phase
in each region.}\label{fig4}
\end{figure}

In conclusion, the changes in the frequency of the mode labeled by
C when temperature is varied from 25 to -240 $^oC$, indicates that
PZT undergoes a phase transition from tetragonal to monoclinic
structure. This result agrees well with those recently reported by
Noheda \textit{et al.}\cite{bn,jag} using high resolution
synchrotron X-ray powder diffraction techniques and dielectric
measurements. We were able to describe this phase transition based
on the Landau-Devonshire phenomenological theory. For this we
introduced as a new order parameter the variable 90$^o\, - \beta$,
where $\beta$ is the monoclinic angle. Using the scarce
experimental data available we estimated the modified free energy
coefficients. Calculations carried out with the modified free
energy confirmed that the monoclinic phase is stable in a thin
region close to MPB.

\vskip0.5truecm\noindent Acknowledgements~~--~~ Financial support
from FUNCAP, CNPq, FAPESP and FINEP is grateful acknowledged. One
of us, A. G. Souza Filho wishes to acknowledge the fellowship
received from FUNCAP.

\end{document}